**Magnetic manipulation of superparamagnetic colloids in droplet-based optical devices**

I. Mattich[1], J. Sendra[2], H. Galinski[2], G. Isapour[3], A.F. Demirörs[1], M. Lattuada[4], S. Schuerle[5], A. R. Studart[1]

[1] Complex Materials, Department of Materials, ETH Zürich, 8093 Zürich, Switzerland

[2] Laboratory for Nanometallurgy, Department of Materials, ETH Zürich, 8093 Zürich, Switzerland

[3] Department of Mechanical Engineering, MIT, USA

[4] Department of Chemistry, University of Fribourg, Switzerland

[5] Department of Health Sciences and Technology, Institute for Translational Medicine, ETH Zürich, 8093 Zürich, Switzerland

**Abstract**

Magnetically assembled superparamagnetic colloids have been exploited as fluid mixers, swimmers and delivery systems in several microscale applications. The encapsulation of such colloids in droplets may open new opportunities to build magnetically controlled displays and optical components. Here, we study the assembly of superparamagnetic colloids inside droplets under rotating magnetic fields and exploit this phenomenon to create functional optical devices. Colloids are encapsulated in monodisperse droplets produced by microfluidics and magnetically assembled into dynamic two-dimensional clusters. Using an optical microscope equipped with a magnetic control setup, we investigate the effect of the magnetic field strength and rotational frequency on the size, stability and dynamics of 2D colloidal clusters inside droplets. Our results show that cluster size and stability depend on the magnetic forces acting on the structure under the externally imposed field. By rotating the cluster in specific orientations, we illustrate how magnetic fields can be used to control the effective refractive index and the transmission of light through the colloid-laden droplets, thus demonstrating the potential of the encapsulated colloids in optical applications.

**Introduction**

Magnetic fields offer a powerful means to assemble or manipulate colloids for a broad range of fields, from microrobotics to medicine, biotechnology and manufacturing. [1-4] In microrobotics, magnetic stimuli have been used to control the locomotion of miniaturized robots [1, 5] and to design substrates for the remote manipulation of small-scale objects. [6-8] Locally induced hyperthermia and directed transport using iron oxide particles are well-known examples of potential applications in medicine, in which a magnetic stimulus is employed as a non-invasive tool for drug delivery and cancer therapy. [2, 9] In manufacturing, magnetic fields have been exploited for the fabrication of composites with controlled orientation of reinforcing particles [3, 10] or the formulation of reversible adhesives. [11] Moreover, analytical biotechnological assays often rely on magnetic fields to recover surface-functionalized colloids in biological and chemical separation processes. [4, 12] In many of these applications, superparamagnetic iron oxide nanoparticles (SPIONs) are used as the magnetically responsive colloids. [9]

Suspensions of superparamagnetic particles exhibit very rich phase behaviour and dynamics when subjected to time-varying magnetic fields. [13, 14] Their main advantage lies on the fact that these particles act as magnetic dipoles exhibiting a single magnetic domain only in the presence of an external magnetic field. This allows for switching the colloidal state of the suspension from a fluid dispersion to hierarchical assemblies of particles interacting through attractive and repulsive dipolar forces. Depending on the concentration of particles and the type of magnetic stimulus applied, superparamagnetic colloids can assemble into chains, two-dimensional clusters or three-dimensional hierarchical networks. [13] Such colloidal structures have been considered for several prospective applications as microfluidic mixers [15], microswimmers and micropumps [16], cargo transporters [17], and artificial ciliated surfaces. [13, 18] Despite these potential applications and our advanced understanding of their magnetic response, the directed assembly of superparamagnetic particles compartmentalized inside droplets remains to be investigated and technologically exploited.

The compartmentalization of colloidal particles inside droplets and capsules is an effective approach to control the release of molecules in delivery systems [19] and to program the brightness of pixels in commercially available electronic books. [20, 21] In electronic paper, the encapsulated particles are manipulated using an external electrical field to change optical properties locally across large areas. The applied electrical field drives the motion of black and white encapsulated particles of opposite charges to different regions of the capsule, thus allowing for electrical control of the local brightness. [20] In nature, pigment granules compartmentalized in chromatophore cells are also used by cephalopods to change color on demand. [22, 23] In this case, color is generated through the displacement of granules via controlled contraction or expansion of encapsulating sacks. [22] Such a strategy has inspired the development of magnetically controlled smart windows. [24] These biological and technological examples demonstrate the potential of controlled colloidal manipulation in compartments as an enticing approach to create novel functionalities. Given the increasing availability of multiferroic materials that can generate magnetic fields using low-power electrical input, [25, 26] the use of magnetic torques and forces to drive and control the assembly of colloids in droplets is an interesting and technologically relevant approach that calls for further scientific research.

Here, we study the assembly and manipulation of superparamagnetic colloids inside droplets driven by a time-varying external magnetic field. The magnetically responsive colloids are encapsulated in water-in-oil droplets through a microfluidic emulsification approach. After encapsulation in monodisperse droplets, the particles are assembled and manipulated using a tunable rotating magnetic field. Next, the response of the colloids to the external field is studied by optical microscopy imaging of multiple droplet arrays. Finally, we demonstrate how superparamagnetic particles in droplets can be potentially exploited as magnetically controlled optical shutters and microlens arrays with tunable focal length.

**Results and Discussion**

The assembly of colloids under rotating magnetic fields is experimentally studied by encapsulating monodisperse colloidal particles inside monodisperse water droplets suspended in a continuous oil phase (Figure 1). The colloids consist of polystyrene particles with an average size of 480 nm loaded with superparamagnetic iron oxide nanoparticles (SPIONs, 10-20 nm) to render them magnetically responsive. Due to the high density of carboxylic acid groups (COOH) on the surface (> 50 µmol/g), the particles become negatively charged in water at neutral and alkaline pHs. 1-decanol is used as the continuous phase, whereas the aqueous droplet is prepared from double deionized water. The concentration of colloidal particles inside the droplet is varied between 0.01 and 0.78 vol% to generate clusters that are large enough for visualization by optical microscopy while keeping the system sufficiently dilute to facilitate assembly.

To study the assembly of the particles into two-dimensional clusters, we apply a time-varying external magnetic field to a monolayer of colloid-laden droplets using two types of magnetic stimuli (Figure 1a-c). In a first approach, an external magnetic field is used to drive the assembly of 2D colloidal clusters. To this end, the applied field is rotated within the plane orthogonal to the viewing direction (xy-plane), thus facilitating visualization of the cluster assembly process. In a second step, the same rotating magnetic field is first used to form the 2D clusters and is then employed to manipulate the colloidal assembly within the droplet. This manipulation is accomplished by applying the rotating field in a plane parallel to the viewing angle and revolving it around the viewing axis to rotate the 2D colloidal cluster. To study the dynamics of the colloidal assembly process and the magnetic manipulation of the resulting clusters, the droplets need to be sufficiently stable against coalescence and coarsening events.

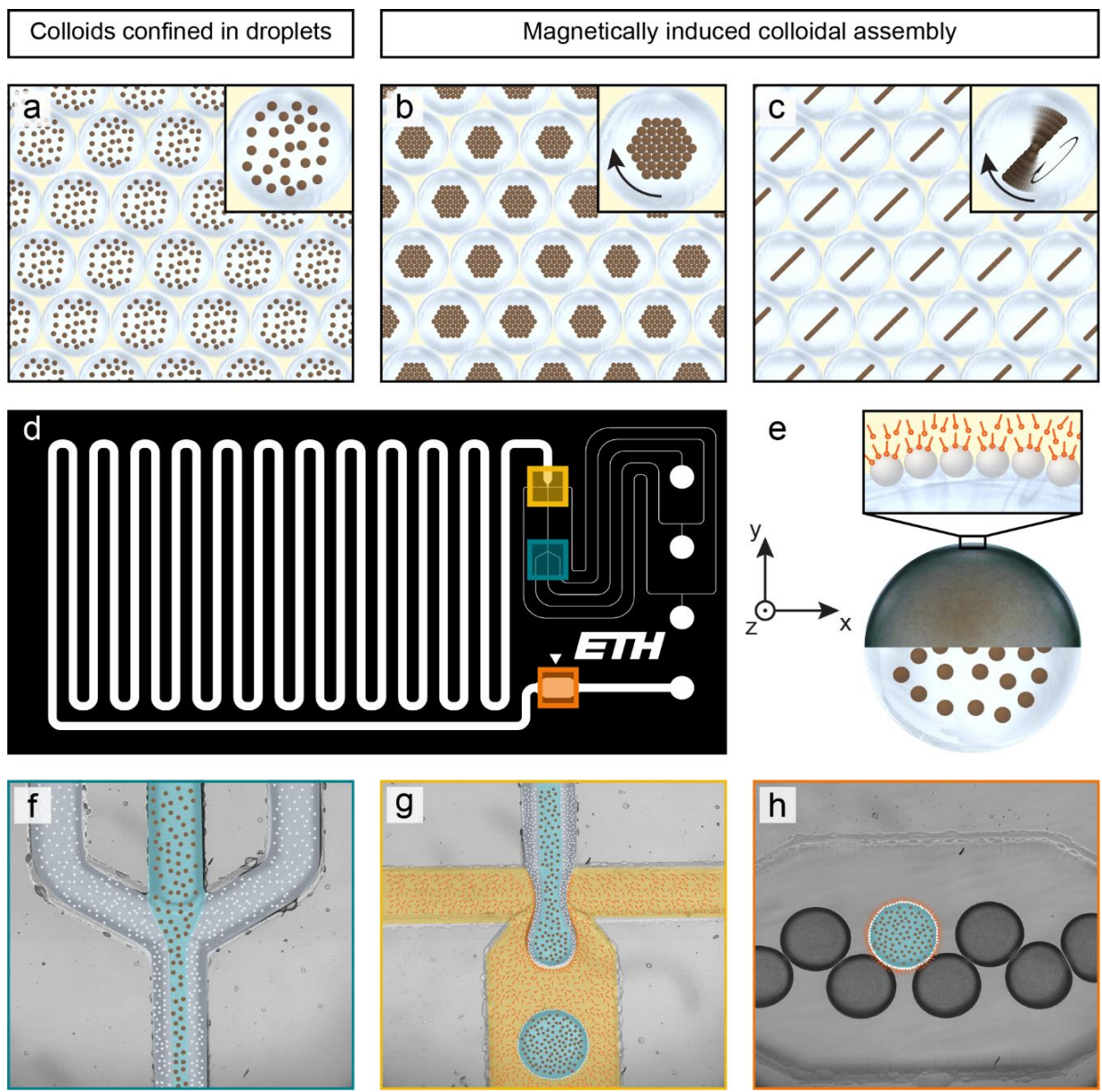

**Figure 1.** Microfluidic encapsulation and magnetically driven assembly of colloidal particles inside droplets. (a) Illustration of colloids loaded with superparamagnetic nanoparticles confined in water-in-oil droplets. When no magnetic field is applied, the colloidal particles undergo random Brownian motion. (b) Schematic of the assembly of colloidal particles into large two-dimensional clusters under a rotating magnetic field with a frequency of 125 Hz applied within the xy-plane. (c) After cluster formation, the axis of rotation of the rotating field can be slowly adjusted until it lies parallel to the y-axis, thus shifting the color of the droplet from dark to translucent. At this point, the rotating field plane is revolved around the z-axis at a certain pre-defined speed to study the mechanical stability of the cluster. (d) Overview of the flow-focusing microfluidic device, highlighting the junctions and chamber where droplets are formed and collected. (e) Cartoon illustrating the stabilization of the water-oil interface by surface-modified silica nanoparticles and surfactant molecules. (f) Magnification of the first junction, indicating the laminar co-flow of the aqueous phases containing the superparamagnetic colloids (center) and the silica nanoparticles (close to walls). (g) Magnification of the second junction, where water-in-oil droplets are formed by flow-induced dripping. (h) Magnification of the observation chamber at the end of the serpentine channel. The droplets are stable and do not coalesce upon physical contact. The particles, fluids and surfactants were false-colored in the images to facilitate visualization.

Monodisperse, stable droplets were generated in a flow-focusing microfluidic device using Pickering emulsions as templates (Figure 1d-h). The Pickering emulsions were formed by adsorbing silica nanoparticles at the water-oil interface. In this microfluidic emulsification approach, water droplets are formed in a continuous oil phase through a flow-induced dripping mechanism. [27] Emulsification occurs by injecting the aqueous inner phase and the oil outer phase through separate input channels of the microfluidic device (Figure 1d). The channels are designed to form an aperture, at which the dripping phenomenon takes place (Figure 1g). The aqueous inner phase consists of a suspension of superparamagnetic colloidal particles in water, whereas 1-decanol is used as the outer oil phase.

To facilitate the Pickering stabilization of the oil-water interface, a second coaxial aqueous phase is added in the device during emulsification. Such liquid phase comprises the silica nanoparticles suspended in water and is injected alongside the innermost aqueous phase in a co-flow configuration. Such an approach enables the delivery of the silica nanoparticles very close to the oil-water interface formed upon dripping. The silica nanoparticles were partially hydrophobized with hexyl amine to favor their adsorption at the oil-water interface. Interfacial adsorption of the silica particles is aided by a silicone-based surfactant (ABIL 90 EM) added to the oil phase.

Droplet microfluidics enabled precise control over the emulsion droplet size, monodispersity, droplet surface coverage, and concentration of encapsulated superparamagnetic particles. By tuning the concentration of surfactant and modified silica nanoparticles, stable Pickering emulsions were obtained at the end of the serpentine channel and outside of the microfluidic device. This high stability was crucial to study the assembly and manipulation of the encapsulated superparamagnetic nanoparticles using an external magnetic field.

Magnetic fields with a rotational frequency of 125 Hz were applied to a monolayer of droplets to study the assembly and manipulation of colloidal clusters (Figure 2). Because of their negative surface charges, the colloids are initially electrostatically stabilized and uniformly dispersed within the droplets (Figure 2a). The use of a magnetic setup with eight electromagnetic coils and five degrees of freedom [28, 29] allowed us to gain full control over the magnetic field applied to the colloidal particles. Comparative experiments with particles suspended in water showed that their encapsulation inside droplets is crucial to prevent cluster-cluster interactions and thus allow for the systematic investigation of the assembly and dynamics of individual 2D clusters. The high monodispersity of the droplets led to crystallization in ordered domains, thereby facilitating the imaging of the colloidal assembly process in multiple droplets simultaneously.

Our experiments revealed that the initially dispersed colloidal particles readily assemble into 2D clusters when subjected to a rotating magnetic flux density ($B$) with amplitude in the range of 6-11 mT and a frequency ($f$) of 125 Hz (Figure 2). The magnitude of the field applied corresponds to a magnetic field strength between 1132 and 2076 A/m. The assembly process is governed by magnetic interactions between the field-induced magnetic dipoles within individual particles. The superparamagnetic nature of the colloids leads to induced dipoles within the plane of the rotating field, which attract each other to initially form small clusters of colloidal particles. Over time, the small clusters merge into a large two-dimensional assembly with one or two layers of particles, depending on the initial colloid concentration.

While the dipole-dipole interactions in the bulk of the cluster cancel each other, the particles positioned at the edge of the cluster experience a net magnetic moment due to the absence of neighboring colloids outside the cluster. [14] The dipolar-dipolar attractive interactions between such particles lead to a line tension ($\lambda$), which is a 2D analogue of the surface tension of three-dimensional matter.

The emergence of a line tension allows us to magnetically control the size and mechanical properties of the assembled cluster. To demonstrate this, we measured the average diameter of the clusters as a function of the applied magnetic field for a constant frequency of 125 Hz (Figure 2c). The experiments reveal a linear decrease in the average cluster size with the applied field. This result can be interpreted by analyzing the effect of the magnetically induced line tension on the structure of the colloidal cluster. The line tension arising from the rotating magnetic field pulls the colloidal particles together into a more compact closely packed structure. This compressive force is counteracted by electrostatic and/or steric repulsive forces between adjacent particles, which eventually establishes a new equilibrium at smaller interparticle distances.

The smaller distance between colloidal particles affects the mechanical properties of the cluster. At closer distances, stronger interactions are expected between the colloidal particles, enhancing the stiffness of the cluster. To gain an insight into the effect of the magnetic field on the cluster stiffness, we call upon simple scaling relations previously proposed. The shear modulus of the cluster ($G'$), taken here as a measure of stiffness, can be approximated by $\lambda/R$. [14] Earlier work has shown that the tension line $\lambda$ is expected to scale with the square of the field amplitude: $\lambda \sim H^2$. [14] Combining these relations, we conclude that the stiffness of the cluster should scale with the ratio $H^2/R$, a quantity that can be directly calculated from our experimental data. By plotting experimental values of $H^2/R$ as a function of $R$, one finds that the stiffness of the cluster should increase with the reduction in cluster size, as illustrated in Figure 2d. Indeed, decreasing the cluster size by a few percent results in a 5-fold increase in the expected stiffness. This indicates the possibility of using magnetic fields to control the stiffness of the 2D colloidal clusters.

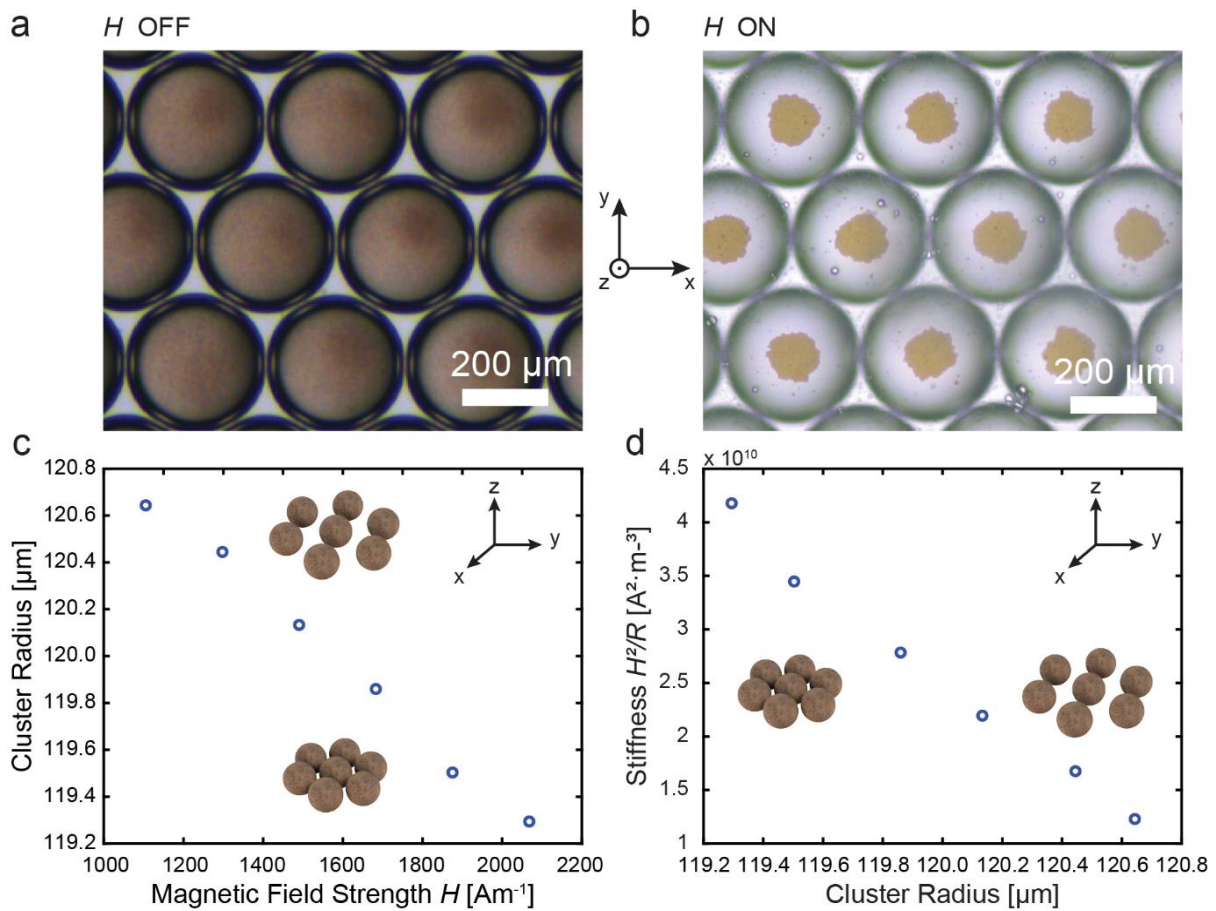

**Figure 2.** Magnetic assembly of superparamagnetic particles into 2D colloidal clusters inside droplets. (a) Optical microscopy image of an array of monodisperse water-in-oil droplets with superparamagnetic particles homogeneously suspended in the aqueous phase. (b) Particles assemble into 2D clusters when a magnetic field rotating at 125 Hz is applied within the xy-plane. (c) Effect of the magnetic field strength ($H$) on the diameter ($2R$) of the assembled colloidal clusters. (d) Impact of the cluster size on the ratio $H^2/R$, which is taken as an indicative value for the stiffness of the colloidal cluster.

In addition to the formation of 2D assemblies with controlled size and stiffness, the rotating field can also be used to manipulate the as-formed colloidal clusters encapsulated within the droplets. Manipulation of the dynamically assembled 2D clusters is possible by moving the plane of rotation while keeping the in-plane frequency active (Figure 3). To demonstrate such magnetic control capabilities, we performed experiments in which 2D clusters are first assembled with a magnetic field rotating in the xy-plane. Afterwards, the axis of rotation of the clusters is turned to become parallel to the y-axis. Finally, their stability is studied by revolving the clusters around the z-axis at increasing angular speeds, starting from 0.1 rad/s (Figure 3a,b). Optical microscopy imaging reveals that the encapsulated clusters can be effectively rotated inside the droplets without losing their two-dimensional morphology. The motion of the cluster into a new orientation changes dramatically the transmittance of light across the droplet and thereby the optical properties of the entire droplet array.

The ability of the cluster to remain stable during magnetic manipulation depends on the applied revolving speed ($\omega$) and the external magnetic field strength ($H$). We experimentally observed that revolving speeds above a critical value ($\omega_c$) lead to extensive fragmentation of the 2D clusters into smaller colloidal assemblies (Figure 3c and Movie S1). To better understand this fragmentation phenomenon, we measured the critical revolving speeds for clusters of different sizes subjected to distinct external rotating magnetic field strengths (Figure 3d). Our results show that the critical revolving speed increases non-linearly with the strength of the rotating field applied ($H$). For a cluster intermediate size of 70 μm, we find that the $\omega_c$ value doubles when the magnetic field increases from 1321 to 2265 A/m. Moreover, smaller clusters can be revolved without fragmentation at a higher velocity compared to their larger counterparts. Under a magnetic field strength of 673 A/m, the threshold revolving speed drops from 3.37 to 1.42 rad/s if the cluster size is enlarged from 50 to 100 μm.

The effect of the magnetic field strength and cluster size on the critical revolving speed ($\omega_c$) can be interpreted by considering the forces acting on the cluster during magnetic manipulation. Earlier studies have shown that a balance between the magnetic torque $T_m$ arising from the applied field and the reactive viscous torque $T_\eta$ exerted by the liquid governs the dynamics of anisotropic assemblies and particles under rotating magnetic fields.[13, 30, 31] For steady-state rotation of our system, a net balance of these forces is achieved such that $T_m + T_\eta = 0$, leading to the following expression:

$$\omega - \omega_c \sin 2\theta = \frac{d\theta}{dt} \quad (1)$$

with

$$\omega_c = \frac{\mu_0 \chi_{cluster}^2}{12\eta_0 \left(f/f_0\right)(1+\chi_{cluster})} \cdot H_0^2 \quad (2)$$

Here, $\theta$ represents the phase lag between the plane of the rotating magnetic field and the cluster's long axis, while $\omega$ is the fixed magnetic field angular speed applied. $\mu_0$ indicates the magnetic permeability of free space, $\chi_{cluster}$ is the effective magnetic susceptibility of the cluster and $H_0$ is the applied magnetic field. $\eta_0$ is the viscosity of the fluid surrounding the cluster and $\left(f/f_0\right)$ represents the Perrin friction factor. The frequency of rotation of the cluster is coupled with the magnetic field rotational frequency when the phase lag is constant over time ($\frac{d\theta}{dt} = 0$). Assuming that inertial and gravitational effects can be neglected, the anisotropic object will follow the rotating magnetic field if its frequency $\omega$ lies below the critical value, $\omega_c$. When this critical condition is surpassed, the rotating object cannot keep up with the speed of the rotating field, resulting in structural instability and loss of synchronous motion.

The theoretical analysis above predicts that the critical angular speed ($\omega_c$) measured in our experiments should scale with the square of the applied magnetic field strength ($H_0$): $\omega_c = kH_0^2$, with $k$ depending on the particle magnetization, cluster diameter and fluid viscosity. We test this prediction by comparing the analytical model derived in Eq.2 to our experimental data (Figure 3d). In this comparison, we estimate the parameter $k$ assuming values for $\mu_0, \chi_{cluster}, \eta_0$, and $f/f_0$ that are known for our system or available in the literature (supporting information). The good agreement observed between experiments and the analytical model suggest that the dynamics of the rotating clusters is captured well by the torque balance proposed in the literature (Figure 3d). The decrease in the $k$ value with increasing cluster size reflects the stronger viscous forces exerted on larger clusters compared to smaller counterparts. This explains the lower critical rotational speeds experimentally needed for the fragmentation of larger clusters. Fragmentation lowers the viscous forces acting on the newly formed smaller clusters, which are stable enough to synchronously rotate with the applied magnetic field (Movie S1). Catastrophic fragmentation events were more difficult to detect for smaller cluster sizes. This may explain the discrepancy for smaller cluster diameters with our analytical model.

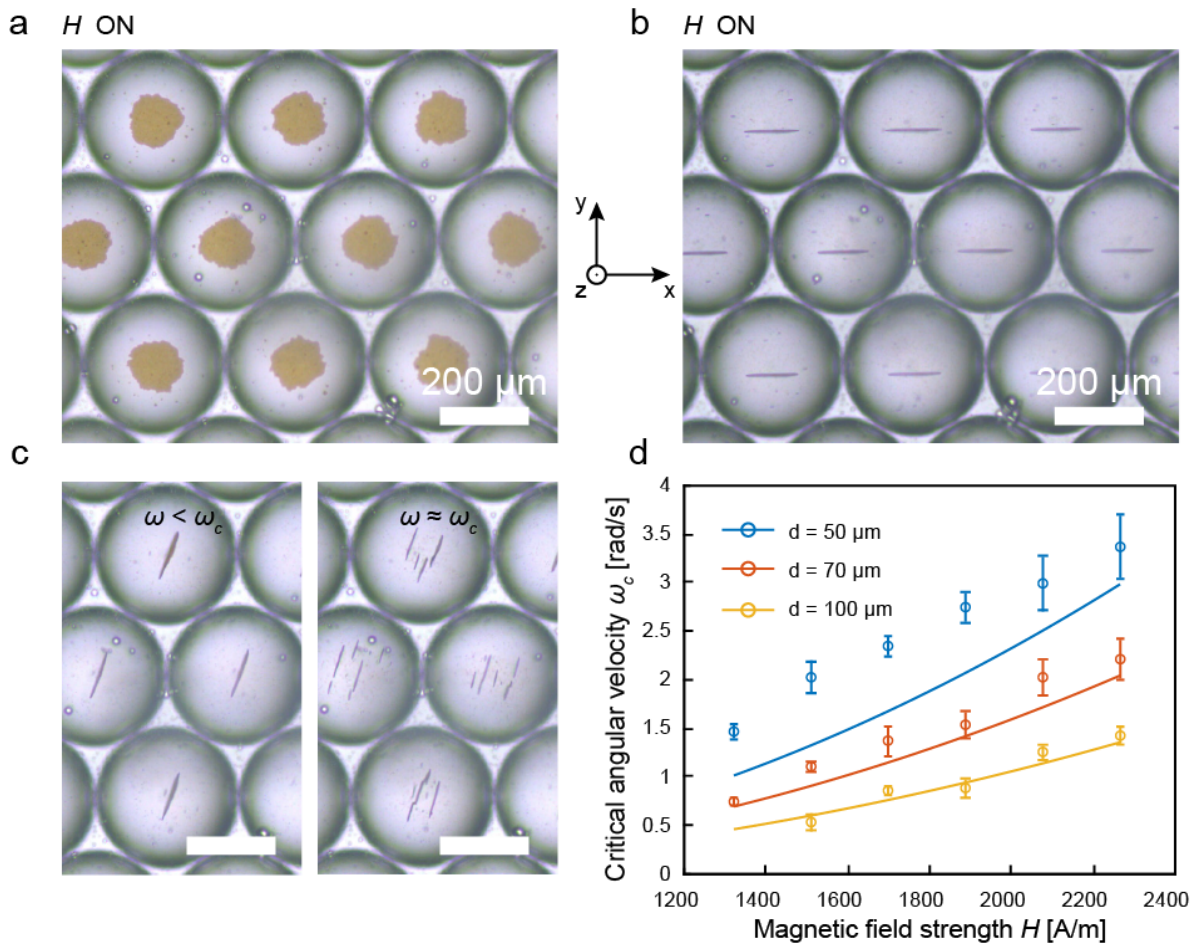

**Figure 3.** Manipulation and fragmentation dynamics of magnetically assembled 2D colloidal clusters. (a) Optical microscopy image depicting an array of encapsulated clusters formed in the presence of a magnetic field rotating within the xy-plane. (b) Reorientation of the clusters shown in (a) achieved by turning the axis of rotation around the x-axis. (c) Fragmentation of the clusters observed when the revolving speed ($\omega$) of the rotating field is increased above the critical value, $\omega_c$. In this experiment, the plane of the rotating magnetic field was revolved around the z-axis. (d) Effect of the magnetic field strength on the critical revolving speed ($\omega_c$) for clusters of different diameters $d$. The continuous lines represent the analytical solution $\omega_c = kH_0^2$, with $k$ values calculated from parameters that are known for our system (Table S1, supporting information).

The formation and manipulation of 2D colloidal clusters under a rotating field opens new opportunities to magnetically control the optical properties of individual droplets. The use of magnetic fields to control the refractive index and light transmittance of the droplets can potentially be exploited for the development of active optical components, such as shutters and microlens arrays. To illustrate this potential, we first study the transmittance of light across monodisperse droplets under magnetic fields and later assemble similar droplets into a polymer matrix to create magnetically controlled soft optical shutters.

In the first demonstrator, a colloid-laden water-in-oil droplet is used as a microlens with magnetically controlled effective refractive index. The effective refractive index is affected by the spatial distribution and assembly of the superparamagnetic colloids inside the droplet. By changing the effective refractive index of the droplet, it is possible to actively control the focal length of the microlens. In contrast to inspiring previous research on droplet-based lenses, [32] the focal length in our demonstrator is actively controlled by the external magnetic field. To quantify this effect, we measured the change in the focal length of a single microlens upon exposure to a magnetic field parallel or perpendicular to the incoming light (Figure 4). For these measurements, a table-top optical setup was built around the electromagnetic coils that apply the external magnetic field (Figure 4d,e). A sample holder was 3D printed to host the monolayer of droplets loaded with superparamagnetic colloids (Figure S1).

For the experiment, a monochromatic collimated laser beam with a wavelength of 532 nm illuminates the sample and the refracted beam intensity profile is mapped around the resulting focal point. To enable the assembly and manipulation of 2D colloidal clusters inside the individual droplets, a magnetic field rotating at 125 Hz was applied following the protocol stated above. Before the magnetic field is applied, the droplet contains a homogeneous dispersion of particles (state 1). The change in focal length of the droplet-based microlens was measured for two magnetically induced conditions: 2D clustering parallel to the input beam (state 2) and 2D clustering perpendicular to the input beam (state 3).

The experimental results show that we can discretely change the focal length depending on the state of the superparamagnetic colloids (Figure 4a-c). Because of the lower refractive index of water compared to the oil, the focal point for all the investigated states is positioned between the light source and the droplet-based microlens. The magnetically induced assembly of dispersed particles into clusters directly affects the experimentally measured focal length. To evaluate this effect, we set as reference the focal length created by droplets with parallel-aligned clusters (state 2, z=0 in Figure 4b) and report the shift in focal length observed when the magnetic field is switched off (state 1, Figure 4a) and when clusters are assembled perpendicular to the incoming light (state 3, Figure 4c). In the reference state (2), the cluster is oriented parallel to the beam and interacts minimally with the incoming light, leading to droplets with optical properties dominated by the aqueous phase. Switching off the field (state 1) leads to thermal randomization of the particles and a shift in focal length of 92 µm. The focal length change reduces to 84 µm, if the cluster is magnetically oriented perpendicular to the input beam (Figure 4b,c).

To establish a quantitative correlation between the change of the microlens focal length and the assembly of superparamagnetic colloids in the droplet, we applied a ray tracing analytical model and finite element simulations to our optical system. Using the transfer matrix method, the analytical model predicts the inverse of the focal length ($1/f$) to depend on the effective refractive index of the droplet, the refractive index of the oil and the radius of the droplet (see SI and Figure S2). For a given oil and droplet size, the analytical model indicates that the focal length of the microlens should decrease continuously with the effective refractive of the droplet, $n_{eff}$ (Figure 4f). To complement this theoretical model, we performed finite element simulations on a representative droplet illuminated with monochromatic collimated light (Figure 4g and Figure S3). The simulations confirm that the droplet diverges the incoming light, leading to a focal point positioned between the droplet and light source, as observed in the experiments (Figure 4h and Figure S3). Changes in the refractive index of the droplet results in a shift of the simulated focal length, in close agreement to the analytical model (Figure 4f). By exploring other materials and droplet types, the simulations provide useful guidelines for the design of the droplet-based microlens (Figure S4).

The simulations and the analytical model suggest that the focal length change observed when the parallel-aligned clusters (state 2) disperse into homogeneously distributed colloids (state 1) can be explained by a change in the effective refractive index of the droplet. Assuming that the refractive index of the droplet with the parallel-oriented cluster is dominated by water ($n_{eff} = 1.330$, red circle in Figure 4f), our simulation predicts that the effective refractive index of the droplet should increase to 1.345 for the homogeneously suspended colloids (dashed line, Figure 4f). While the magnetic nature of the suspended colloids prevents us from estimating the effective refractive index of the homogeneous droplet, the higher refractive index of polystyrene ($n = 1.58$) compared to water makes our physical interpretation a reasonable qualitative explanation for the experimental observations. Because of the heterogeneous nature of droplets with perpendicularly aligned clusters, we expect the focal length change induced by this configuration to arise from the complex interactions of the beam with the particles within the oriented cluster.

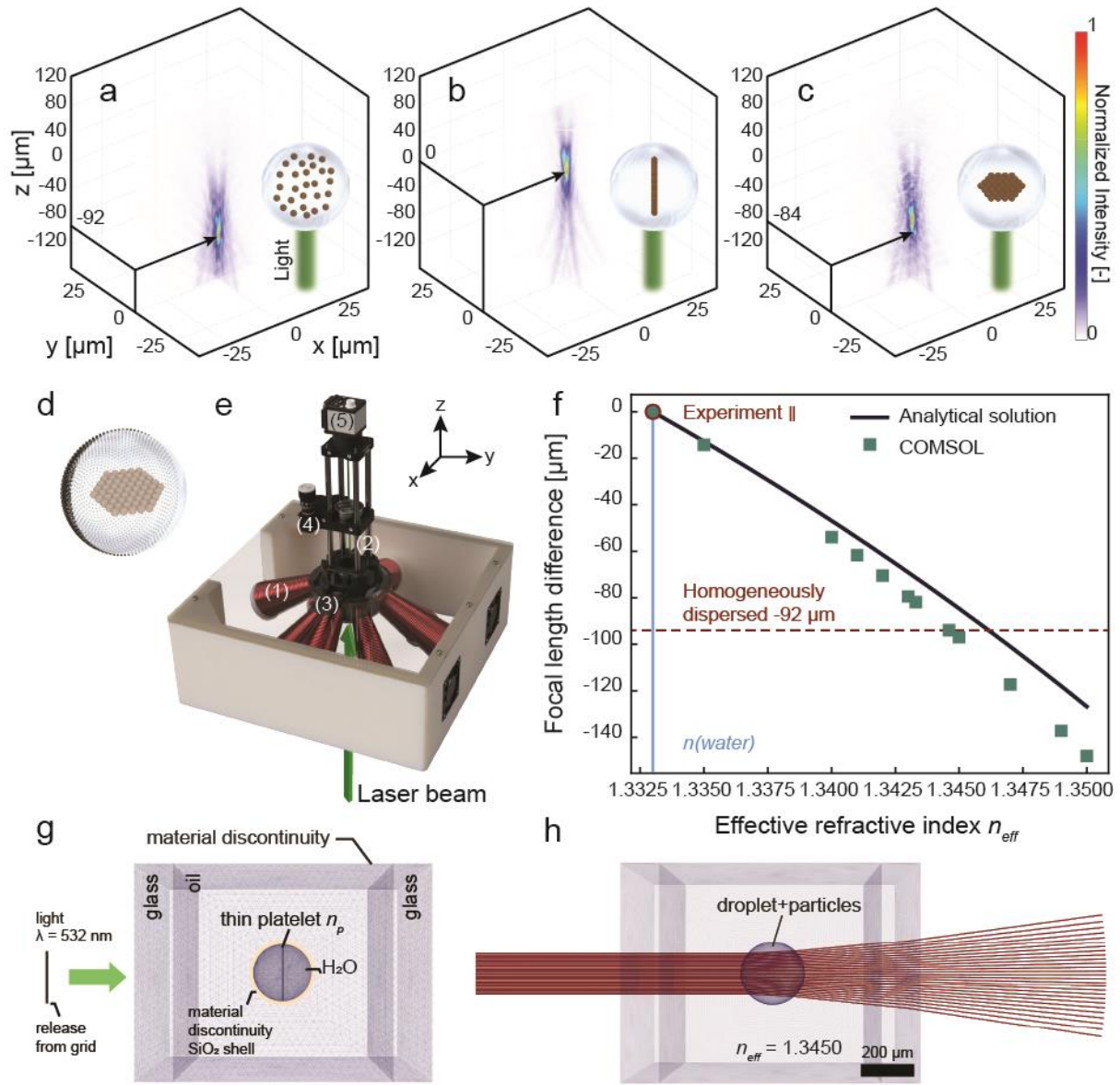

**Figure 4.** Active microlenses comprising water-in-oil droplets loaded with magnetically responsive superparamagnetic colloids. (a-c) Maps of the light intensity spatial distribution resulting from the refraction by droplets containing colloids in different configurations: (a) homogeneously dispersed superparamagnetic nanoparticles (no magnetic field), (b) colloidal cluster oriented with its plane parallel to the laser beam, (c) colloidal cluster oriented with its plane perpendicular to the laser beam. (d) Cartoon showing a Pickering stabilized droplet containing a monolayer cluster of magnetically responsive nanoparticles. Only half of the surface of the droplet is covered with interfacially adsorbed colloids to better visualize the encapsulated colloidal cluster. (e) Rendering of the optical setup used to measure the change in focal length of individual droplet-based lenses under the action of a magnetic field. The individual components of the setup are: (1) magnetic coils, (2) laser beam, (3) sample, (4) UV lens mounted on a z-axis translational lens mount and (5) detector. (f) Effect of the effective refractive index of the droplet on the change in focal length using a colloid-free water droplet as reference. Experimental data is shown in red, green triangles indicate results from ray tracing simulations and the black line is the analytical solution. (g) Illustration of the three-dimensional ray optics model used for the finite element simulations, including boundary and ray release conditions. (h) Simulated ray trajectories ($\lambda$=532 nm) for a homogeneously dispersed suspension of particles with effective refractive index of $n_{eff}$ =1.3450, corresponding to a change in focal length of -92 μm as measured in the experiment.

In addition to microlens arrays, the colloid-laden droplets can also be exploited as magnetically controlled optical shutters. We provide an example of such an application by preparing a demonstrator that consists of a polymer monolith containing a monolayer of water droplets (Figure 5a,b). The monolith is made by using an oil continuous phase that can be polymerized after the assembly of the microfluidic droplets (Figure S5). In this case, a high concentration of superparamagnetic particles of 0.78 vol% is used in order to amplify the difference in light transmission through the monolith when the shutter is in the ON and OFF states. A rotating magnetic field is applied to assemble the particles into anisotropic structures aligned either perpendicular or parallel to the incoming light, thus switching the shutter ON or OFF, respectively. By using a low frequency up to 5 Hz, we expect the particles to assemble into oriented clusters under the applied field.

The performance of the optical shutter was assessed by measuring the evolution of the transmitted light while the plane of the rotating magnetic field was periodically changed between the parallel and perpendicular orientations. Experiments were carried out under an optical microscope using magnetic flux density magnitudes in the range 1 – 20 mT. The transmitted light intensity was obtained directly from optical microscopy images using image analysis software (Figure 5c,d).

Our results reveal that the transmitted light intensity can change up to two-fold when the plane of the rotating magnetic field is switched between the perpendicular and parallel orientations (Figure 5c). Images of the droplet when exposed to the parallel field orientation confirm the assembly of the particles into multiple anisotropic structures, the alignment of which favors light transmission. When the field is changed to the perpendicular orientation, the anisotropic colloidal structures are no longer visible due to complete blockage of the incoming light. The oscillations in transmitted light observed in the OFF state match the frequency of the applied magnetic field (Figure 5d and Figure S6) and probably result from the oblate geometry of the colloidal clusters. This suggests that the magnetically assembled structures oscillate around the plane of the magnetic field despite being locked in the imposed orientation.

The timescale of the light intensity changes depends on the switching direction and on the magnitude of the magnetic field. Switching the shutter from OFF to ON (darkening effect) happens at a speed that depends on the magnetic field applied. The time it takes to switch from the maximum to the minimum intensity value for field flux densities of 5, 10, 15 and 20 mT are 0.07, 0.15, 0.25 and 0.25 seconds, respectively. Larger applied magnetic fields lead to higher maximum intensity values, resulting in longer time intervals for them to return to their baseline transmittance levels. The effect of the magnetic field strength on the switching time is reversed when we consider the switching from the ON to the OFF states (whitening effect). In this case, the duration necessary to increase the intensity to a certain fixed level decreases with the magnitude of the magnetic field. When a magnetic flux density of 10 mT is applied, it takes 2 seconds to increase the light transmittance above 20%. However, when a field amplitude of 20 mT is used, the same intensity value can be attained in only 0.45 seconds. The strongest whitening effect leads to a 30% rise in light transmittance from the baseline. This is achieved through the application of a magnetic flux density of 20 mT and a rotational frequency of 1 Hz (see SI and Figure S6). The long timescales needed to transition from the ON to the OFF state might result from the relatively slow process of merger of small initial particle chains into larger anisotropic structures. These experimental results provide insights into the several parameters governing the dynamics of the magnetic assembly process. Future research should be dedicated to further explore this parameter space and thus tune the switching timescales to meet technological demands.

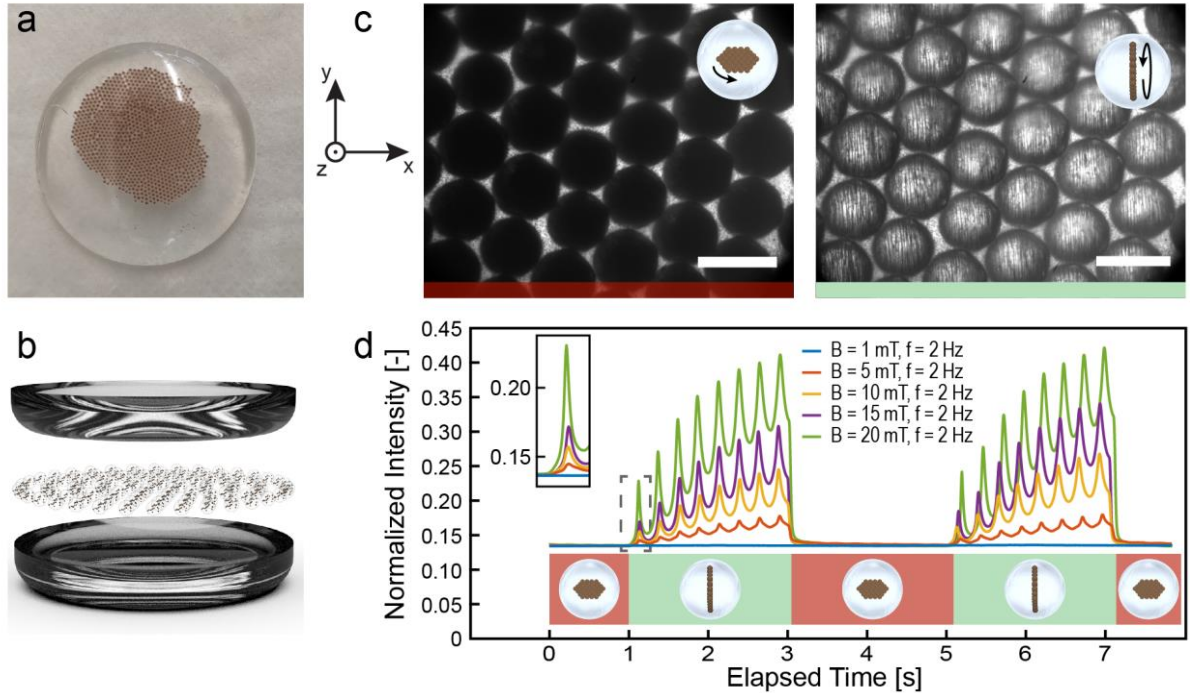

**Figure 5.** Droplet-based optical shutter controlled by magnetic fields. (a) Photograph of a polymer monolith containing colloid-laden water droplets. (b) Rendered sketch showing the manufacturing steps of the optical shutter. (c) Optical microscopy images displaying the color of the droplets when the shutter is ON (left) and OFF (right). A flux field of 20 mT is applied for the OFF and ON states. Scale bar is 500 µm. (d) Intensity of light transmitted across the droplet array when the shutter is subjected to consecutive ON/OFF cycles. The ON state is reached when the magnetic field is rotated within the xy-plane, whereas the OFF state is achieved when the applied field rotates within the yz-plane. The experimental data depict the effect of the magnetic flux density at a fixed frequency of 2 Hz.

**Conclusions**

Superparamagnetic colloids inside droplets can be harnessed to create droplet-based microlenses and optical shutters driven by low-magnitude magnetic fields (6-12 mT). In the presence of a rotating magnetic field, the colloids assemble into tunable two-dimensional clusters driven by the line tension arising from the alignment of magnetic dipoles at the edge of the cluster. The cluster comprises one layer of superparamagnetic colloids and can easily vary between 50 and 250 $\mu$m in diameter, depending on the droplet size and the initial particle concentration in the droplet. Increasing the applied magnetic field strength enhances the line tension, resulting in smaller and stiffer clusters. Clusters can be manipulated by revolving the applied rotating field below a critical frequency. Such a critical frequency scales with the square of the applied field, as predicted by a balance of viscous and magnetic torques acting on the cluster. The ability to assemble and manipulate the 2D colloidal clusters allows one to change the effective refractive index of the droplet and thereby shift the focal length of magnetically controlled droplet-based microlenses. Alternatively, the assembly of colloids into clusters can be used to change the transmittance of light through the droplet, thus leading to magnetically driven optical shutters. In addition to these proof-of-concept demonstrators, the proposed encapsulated colloids may find potential applications in bioinspired color-changing displays, optical devices for telecommunications, camouflage skins, and magnetically writable boards.

**Acknowledgements**

The financial support from ETH Zürich and from the Swiss National Science Foundation through the National Center of Competence in Research (NCCR) for Bio-Inspired Materials is gratefully acknowledged. The authors also thank Dr. Tom Valentin and Dr. Nima Mirkhani for the introduction to the magnetic system MFG-100 from MagnetobotiX®. The glass devices were fabricated at the ETH center for micro-and nanoscience, FIRST.

Conflict of interest: S. Schuerle is co-founder and Member of the Board of MagnebotiX AG.

**Materials and Methods**

Manufacturing of microfluidic devices

Glass microfluidic devices were fabricated using wet etching techniques previously reported in the literature. [33] Briefly, two 1.0 mm thick borosilicate glass wafers (Borofloat 33 Schott) were annealed for 4 h at 580 °C to remove any internal stresses caused by the polishing during production. 50 nm Cr and 50 nm Au thick layers were deposited on the glass substrates using electron beam physical vapor deposition. The wafers were spin-coated with a negative photoresist (AZ10XT 520 cp, thickness of 8 μm) and templated with a sequence of photolithography steps. Both wafers were chemically etched in 40% HF at an etching rate of 3.2 μm/min. The flow-focusing device was etched to a depth of 120 μm, while for the open step emulsification device channels 100 μm deep were produced. Following the etching step, the mask was removed in succession using acetone, isopropanol, gold etchant, and chrome etchant. The wafers were diced into 15 × 30 mm$^2$ (flow-focusing) or 15 × 50 mm$^2$ (step emulsification) chips and 0.7 mm inlets were drilled using a diamond-coated drill bit. Next, the wafers were meticulously cleaned with acetone and isopropanol and subjected to two surface activation steps, first in piranha (1:1, sulfuric acid:hydrogen peroxide) and later in RCA1 (1:1:5, ammonium hydroxide:hydrogen peroxide:water) solutions. After such treatments, the wafers were flushed with water and manually aligned while still wet. Weakly bonded chips were obtained after drying for several hours. A stronger connection was achieved by thermal bonding of two symmetric glass wafers for 4 h in a furnace at a temperature of 620 °C, which is slightly above the glass transition temperature ($T_g$) of borosilicate glass. To ensure good contact between the two wafers, weights were placed on top of each chip, resulting in an applied pressure of 22 kPa (15 × 30 mm$^2$) and 26 kPa (15 × 50 mm$^2$). The success of the bonding step was controlled by visual inspection under an optical microscope.

Surface treatment of microfluidic channels

The microfluidic device was connected to plastic syringes (BD Luer-Lok™ Syringe), mounted on displacement-controlled pumps (Pump 33 DDS, Harvard Apparatus), using PTFE tubing (outer diameter of 1.6 mm, inner diameter of 0.8 mm, Bohlender) and a 4-way linear connector (Dolomite Microfluidics).

To form water-in-water-in-oil or water-in-oil emulsion droplets, the channels of the microfluidic devices need to be surface treated to become hydrophobic. To prepare for such surface treatment, the devices were cleaned in a furnace (model LT 5/11/B410, Nabertherm) at 550 °C for 4 h, then flushed with 1 M NaOH and rinsed with water. The functionalization was performed by flowing a toluene solution containing 5 vol% octadecyltrimethoxysilane (ODTMS, Acros) and 0.5 vol% butyl amine (Acros) for 2 h at a flow rate of 100 µL/h.

Microfluidic emulsification

Water droplets loaded with magnetic nanoparticles were prepared by microfluidic emulsification. Monodisperse magnetic responsive polystyrene nanoparticles were purchased from MicroParticles GmbH (PS-MAG-COOH). The as-received particles show a mean diameter of 536 nm and contain COOH functional groups on the surface. The polystyrene is loaded with an iron oxide content > 30 wt% and the surface functionalized with carboxyl groups. The particles are supplied as a suspension with 1.96 wt% solid content. Such dispersion was diluted with Milli-Q water at different concentrations ranging from 0.01 wt% up to 0.78 wt% and used as dispersed phase of the emulsions.

Unless mentioned otherwise, the water droplets were covered with silica nanoparticles to form highly stable Pickering emulsions. The silica nanoparticles were suspended in water and delivered as a co-flowing aqueous middle phase during the emulsification process (Figure 2, main text). The suspensions contained 10 vol% silica nanoparticles with a mean diameter 120 nm (SNOWTEX ZL, Nissan Chemicals). The particles were partially hydrophobized with hexyl amine (Acros) to a calculated SiOR surface density of 3.22 nm$^{-1}$ before dilution to a concentration of 1.5 vol% to be used as the middle phase. For all the experiments, the flow rates of the dispersed and middle phases were set to 600 µl/h and 300 µl/h, respectively. This leads to a theoretical surface coverage of around three monolayers. The long serpentine channel in the chip was designed to give the silica nanoparticles enough time to diffuse and adsorb at the water/oil interface.

The continuous phase of the emulsions was formulated depending on the targeted experiments or demonstrations. For the experiments on the assembly and fragmentation dynamics of 2D colloidal clusters, the non-ionic surfactant cetyl PEG/PPG-10/1 dimethicone (ABIL 90 EM, kindly provided by Evonik) was dissolved in 1-decanol at a concentration of 2 wt% and used as continuous phase. This non-aqueous surfactant solution was also employed for the demonstration of magnetically controlled active lenses. For the magnetic shutter experiments, 2 wt% polyglycerol polyricinoleate (PGPR 4150, kindly provided by Palsgaard) was used as surfactant in the oil phase, which consisted of isodecyl methacrylate (IDMA, Sigma-Aldrich) and 1,6-hexanediol dimethacrylate (HDDMA, TCI) with a weight ratio of 7:3 and 1 wt% 2,4,6-trimethylbenzoyldiphenyl phosphine oxide (TPO, Sigma-Aldrich) as photoinitiator. Because the continuous phase was polymerized right after emulsification, no Pickering stabilization with silica nanoparticles was employed for the fabrication of the magnetic shutters. For this set of experiments, a step emulsification glass microfluidic device with 100 µm step height was employed.

Magnetic alignment setup

The assembly and manipulation of the magnetic responsive nanoparticles was performed using the magnetic system MFG-100 from MagnetobotiX®, which generates arbitrary magnetic flux densities up to 20 mT at frequencies up to 2 kHz. As described in previous work, [34] the arrangement of eight electromagnets into two sets allows for unrestrained rotational freedom of the magnetic field. This is accomplished by the superposition of multiple magnetic fields, which creates a homogeneous magnetic field within a spherical workspace with a diameter of approximately 10 mm. Imaging of the colloidal manipulation was performed using an inverted optical microscope (DM IL LED, Leica) equipped with a Leica DFC 295 camera.

Magnetic assembly and manipulation

Monodisperse droplets containing defined concentrations of magnetic nanoparticles were prepared to study the assembly and fragmentation dynamics of two-dimensional superparamagnetic colloidal clusters. To investigate the role of the magnetic field density and rotational frequency over the size of the assembled colloidal clusters, a monolayer of monodispersed droplets containing magnetic responsive particles was deposited on a microscope well slide. The glass surface of the slide was modified to be hydrophobic using a solution containing 5 vol% octadecyltrimethoxysilane and 0.5 vol% butyl amine in toluene. A cover slip was placed on top of the slide to seal the well and reduce the evaporation of 1-decanol. The colloids were assembled into a single cluster using the manual controls of the magnetic setup (pitch, yaw, roll) at a fixed magnetic flux density of 10 mT and a rotational frequency of 30 Hz. Once shaped, a rotational magnetic field was applied within the xy-plane to manipulate the colloidal clusters (Figure 1, main text). The magnetic flux density ranged from 6 to 12 mT with a rotational frequency of 125 Hz and a time step between pulses of 2 ms. The concentration of particles in the droplets was varied from 0.02 to 0.078 wt%. For the fragmentation analysis, clusters with three different approximate diameters were studied: 50, 70 and 100 µm. To facilitate visualization, the clusters were rotated around the x-axis, so that the edge of the clusters pointed in the direction of the observation field. To study the stability of the 2D colloidal assembly, the rotational plane of the clusters was revolved around the z-axis at angular velocities increasing from 0.314 to 2.93 rad/s. For angular velocities between 0.3 rad/s and 1.15 rad/s the increment was set to 0.1 rad/s, while from 1.15 rad/s to 2.93 rad/s the velocity was increased by 0.05 rad/s every revolution. The critical angular velocity at which fragmentation took place was recorder for each cluster inside the droplets.

Active optical elements

The active microlenses array was manufactured by collecting a monolayer of monodisperse droplets on a microscope cover slip (#1.5, $n = 1.523$, [35] VWR), which was previously glued on a 3D printed holder (Figure S1). A 1 mm spacer was placed around the emulsion and covered with a second cover slip (#1.5, $n = 1.523$ (Hibbs A.R., 2006), VWR) to remove the liquid meniscus and to reduce evaporation. The droplets were 271 µm in diameter (coefficient of variation, CV = 1.5%) and contained 0.1061 wt% of magnetic particles. The effective refractive index of the magnetic particles and of the homogeneous particle suspension was calculated using a simple mixing rule. The profile of the light transmitted through the droplets was analyzed using an optical setup comprising a collimated laser beam with a wavelength of 532 nm, a fused silica lens (focal length 50 mm, Thorlabs) mounted on a z-axis translational lens mount (SM1ZA, Thorlabs), a 20x magnification objective (Olympus) and a CMOS camera (DCC3260M, Thorlabs). The setup was aligned using 30 mm cage components (Thorlabs) and a 3D printed connector attached to the magnetic setup.

The magnetic shutter system was fabricated by depositing a monolayer of droplets on a layer of polymerized continuous phase inside a 3D printed substrate (Figure S5). The sample was photopolymerized using a UV light source (Omnicure Series 1000, wavelength range 320-500 nm) with an irradiance of 93 mW/cm$^2$ in a $N_2$ atmosphere for 10 minutes. The droplets were 401.7 µm in diameter (CV 8.3%) and contained 1.96 wt% of magnetic particles. To investigate the change in transmittance of the droplets under an external trigger, rotating magnetic flux densities in the range of 1 – 20 mT and frequencies between 1 and 5 Hz were applied. A sequence of rotational magnetic fields was applied, first revolving on the xy-plane for 2 s and later on the yz-plane for 2 s. The transmittance intensity was measured as a function of time to quantify the response of the magnetic shutter.

**Supporting Information**

**Magnetic manipulation of superparamagnetic colloids in droplet-based optical devices**

I. Mattich[1], J. Sendra[2], H. Galinski[2], G. Isapour[3], A.F. Demirörs[1], M. Lattuada[4], S. Schuerle[5], A. R. Studart[1]

[1] Complex Materials, Department of Materials, ETH Zürich, 8093 Zürich, Switzerland

[2] Laboratory for Nanometallurgy, Department of Materials, ETH Zürich, 8093 Zürich, Switzerland

[3] Department of Mechanical Engineering, MIT, USA

[4] Department of Chemistry, University of Fribourg, Switzerland

[5] Department of Health Sciences and Technology, Institute for Translational Medicine, ETH Zürich, 8093 Zürich, Switzerland

**Content:**

Supporting Text

Supporting Movies

Supporting Figures

**Supporting Text**

Critical frequency of the rotating magnetic field

The revolving speed of the magnetic field above which the clusters undergo fragmentation can be estimated by balancing the magnetic and viscous torques exerted on the assembled cluster. On the basis of previous work, [30, 31] we expect this critical revolving speed ($\omega_c$) to be given by the following equation (see also main text):

$$\omega_c = \frac{\mu_0 \chi_{cluster}^2}{12\eta_0 (f/f_0)(1+\chi_{cluster})} \cdot H_0^2 \quad (S1)$$

where $\mu_0$ is the magnetic permeability of free space ($1.257 \times 10^{-6}\ N/A^2$), $\chi_{cluster}$ is the effective magnetic susceptibility of the cluster, $H_0$ is the magnitude of the applied magnetic field strength, $\eta_0$ is the viscosity of the fluid surrounding the cluster, and $f/f_0$ is the Perrin friction factor. The parameters governing the effect of the magnetic field on the critical revolving speed can be summarized by the term $k$:

$$k = \frac{\mu_0 \chi_{cluster}^2}{12\eta_0 (f/f_0)(1+\chi_{cluster})} \quad (S2)$$

To theoretically predict the dependence of $\omega_c$ on the applied magnetic field strength ($H_0$), we calculate the pre-factor, $k$, for clusters of different sizes and compare the predictions with the experimental results (Figure 3d). The parameters used to calculate $k$ were obtained directly from the literature or were estimated based on other variables of our system (Table S1).

The magnetic susceptibility of the cluster depends on the susceptibility of the single magnetite nanoparticles, $\chi_{magnetite}$, and their volume fraction in the polystyrene-magnetite nanocomposite colloid $\Phi_{magnetite}$: $\chi_{cluster} = \chi_{magnetite} \cdot \Phi_{magnetite}$. The Perrin friction factor for equatorial rotation of an oblate spheroid along one of the long axes is given by:

$$f/f_0 = \frac{\frac{4}{3}\left(\frac{1}{p^2}-p^2\right)}{2-\frac{S}{p^2}}, \quad (S3)$$

with the Perrin S factor for oblate spheroids $S \stackrel{\text{def}}{=} \frac{2\tan^{-1}\xi}{\xi}$, $\xi \stackrel{\text{def}}{=} \frac{\sqrt{|p^2-1|}}{p}$ and axial ratio $p = \frac{h}{R_c}$, where $R_c$ and $h$ are the radius and the thickness of the cluster, respectively. Our experiments show that the clusters comprise a single layer of superparamagnetic nanoparticles. Therefore, we assume the thickness of the cluster to be equal to two times the radius of the magnetite nanoparticles ($2a$). The radius of the clusters depends on the concentration of superparamagnetic colloids in the droplet (Table S1).

Except for the smallest cluster size, the predictions from the above theoretical model are in reasonable agreement with the experimental data obtained for the critical revolving speed ($\omega_c$) as a function of the applied magnetic field for different cluster sizes (Figure 3d). The theoretical underestimation of the $\omega_c$ values for the smallest cluster size suggests that our assumptions are not valid for this experimental condition.

**Table S1**: Parameters used in the calculation of the critical frequency needed for cluster fragmentation under a revolving magnetic field (Figure 3d).

| Sample | Colloidal concentration wt% | Cluster radius $R_c$ [$\mu m$] | Perrin factor $f/f_0$ | Nanoparticle magnetic susceptibility $\chi_{magnetite}$ | Magnetite content vol% | Cluster magnetic susceptibility $\chi_{cluster}$ | Fluid viscosity $\eta_0$ [$Pa \cdot s$] |
|---|---|---|---|---|---|---|---|
| 1 | 0.020 | 25.03 | 44.26 | 20 | 15.32 | 3.22 | 0.01 |
| 2 | 0.039 | 38.45 | 67.99 | 20 | 15.32 | 3.22 | 0.01 |
| 3 | 0.078 | 48.69 | 86.11 | 20 | 15.32 | 3.22 | 0.01 |

Calculation of focal length using analytical model

The optical response of the droplet-based microlens was evaluated using an analytical ray tracing model. In this model, the incoming collimated light interacts with different optical elements of the active lens system, namely two glass slides, the oil medium and the water droplet with the magnetic nanoparticles (Figure S2). The parallel ray originated from the collimated beam (shown in red in Figure S2) goes through the first glass slide without any deviation before encountering the microlens. From there, it goes through four interfaces $C_1$, $C_2$, $I_{og}$ and $I_{ga}$ and three different media $S_w$, $S_o$ and $S_g$. As such, the resulting ray tracing transfer matrix is described as follows:

$$M = I_{ga}S_g I_{og} S_o C_2 S_w C_1 = \begin{bmatrix} A & B \\ C & D \end{bmatrix} \quad (S4)$$

where the propagators through water (inside droplet), oil and glass are defined as:

$$S_w = \begin{bmatrix} 1 & t \\ 0 & 1 \end{bmatrix} \quad (S5)$$

$$S_o = \begin{bmatrix} 1 & d_o \\ 0 & 1 \end{bmatrix} \quad (S6)$$

$$S_g = \begin{bmatrix} 1 & d_g \\ 0 & 1 \end{bmatrix} \quad (S7)$$

Here, $t = 2R$ is the thickness of the microlens, $R$ is the radius of the droplet, $d_0$ is the distance between the droplet surface and the glass slide and $d_g$ is the thickness of the glass slide (Figure S2).

The refractions at the oil-glass and glass-air interface are defined as:

$$I_{og} = \begin{bmatrix} 1 & 0 \\ 0 & \frac{n_o}{n_g} \end{bmatrix} \quad \text{(S8)}$$

$$I_{ga} = \begin{bmatrix} 1 & 0 \\ 0 & n_g \end{bmatrix} \quad \text{(S9)}$$

The refraction at the microlens curved interfaces (here the sign convention has already been applied such that R>0) is described as:

$$C_1 = \begin{bmatrix} 1 & 0 \\ -\frac{n_w - n_o}{R n_w} & \frac{n_o}{n_w} \end{bmatrix} \quad \text{(S10)}$$

$$C_2 = \begin{bmatrix} 1 & 0 \\ -\frac{n_w - n_o}{R n_o} & \frac{n_w}{n_o} \end{bmatrix} \quad \text{(S11)}$$

From the ray transfer matrix, the effective focal length from the principal plane to the focal plane will be $-1/C$. To better compare with the ray tracing simulations, the focal length from the focal plane to the center of the microlens is calculated assuming a small angle approximation such that

$$f_{eff} = \frac{A}{C} + d_g + d_o + R \quad \text{(S12)}$$

Calculating the coefficients of the transfer matrix $M$, one reaches the following expressions:

$$A = 1 - \frac{\alpha t}{n_w} - \frac{1}{f_l}\left(d_o + d_g \frac{n_o}{n_g}\right) \quad \text{(S13)}$$

$$C = -\frac{n_o}{f_l} \quad \text{(S14)}$$

where $\alpha = \frac{n_w - n_o}{R}$ and $f_l$ is the focal length of the microlens, therefore

$$-\frac{1}{f_l} = \frac{\alpha}{n_o}\left(-2 + \frac{\alpha t}{n_w}\right) \quad \text{(S15)}$$

Substituting $n_w$ for an effective refractive index $n_{eff}$ yields the effective focal length in the case for a water droplet loaded with magnetic nanoparticles ($f_{eff}$).

Ray optics simulations

The propagation of light through the microlens was also modelled using the Ray Optics Module of COMSOL Multiphysics. These numerical simulations were not intended to replicate the experimental results one-to-one, but to derive an intuitive physical picture of the light-matter interactions and test the validity of the analytical model introduced above.

In the simulations, a set of rays is propagated through a three-dimensional microlens to mimic the experimental setup introduced in the manuscript (Figure and Figure S3). To add the conformal $SiO_2$ shell ($n_{SiO2-NP} = 1.47$) to the water droplet, a material discontinuity condition on the surface of the water droplet is imposed. The focal length of the system is determined by calculating the intercept of the rays with the x-axis. Here, we calculate the change in focal length as a function of orientation and refractive index for a 4 monolayer (2 µm) thick colloidal cluster. The cluster is assumed to have an ellipsoidal shape with two long semi-axes being 123 µm long and the short semi-axis being 1 µm thick.

While the focal length of the concave microlens is unchanged when the colloidal cluster is oriented parallel to the light rays, the focal length decreases when the cluster is oriented perpendicular. This leads to a defocus. By switching between these two states, the focal length can be altered by 25%. In the absence of an external magnetic field, the colloidal particles are homogeneously distributed within the water droplet. As such, the refractive index of the droplet loaded with colloidal particles can be treated with the effective medium approximation.

In agreement with our analytical ray tracing model, the focal length obtained from the simulation decreases with increasing effective refractive index (Figure 4f). This validation is an important result, as we can then predict the focal length and focal power of the microlens with different droplet configurations (Figure S4).

**Supporting Movies**

Movie S1: Two-dimensional colloidal clusters undergoing fragmentation upon increase of the revolving frequency of the applied rotating magnetic field. Experimental parameters: cluster diameter, 70 µm; particle concentration, 0.039 wt%; magnetic flux density, 12 mT; rotating frequency, 125 Hz; revolving frequency range, 12 deg/s – 180 deg/s with a frequency increase each turn of 6 deg/s.

Movie S2: Magnetically driven optical shutter made from arrays of droplets containing superparamagnetic colloids. Experimental parameters: magnetic flux density, 10 mT; rotating frequency, 2 Hz; particle concentration, 1.96 wt%.

**Supporting Figures**

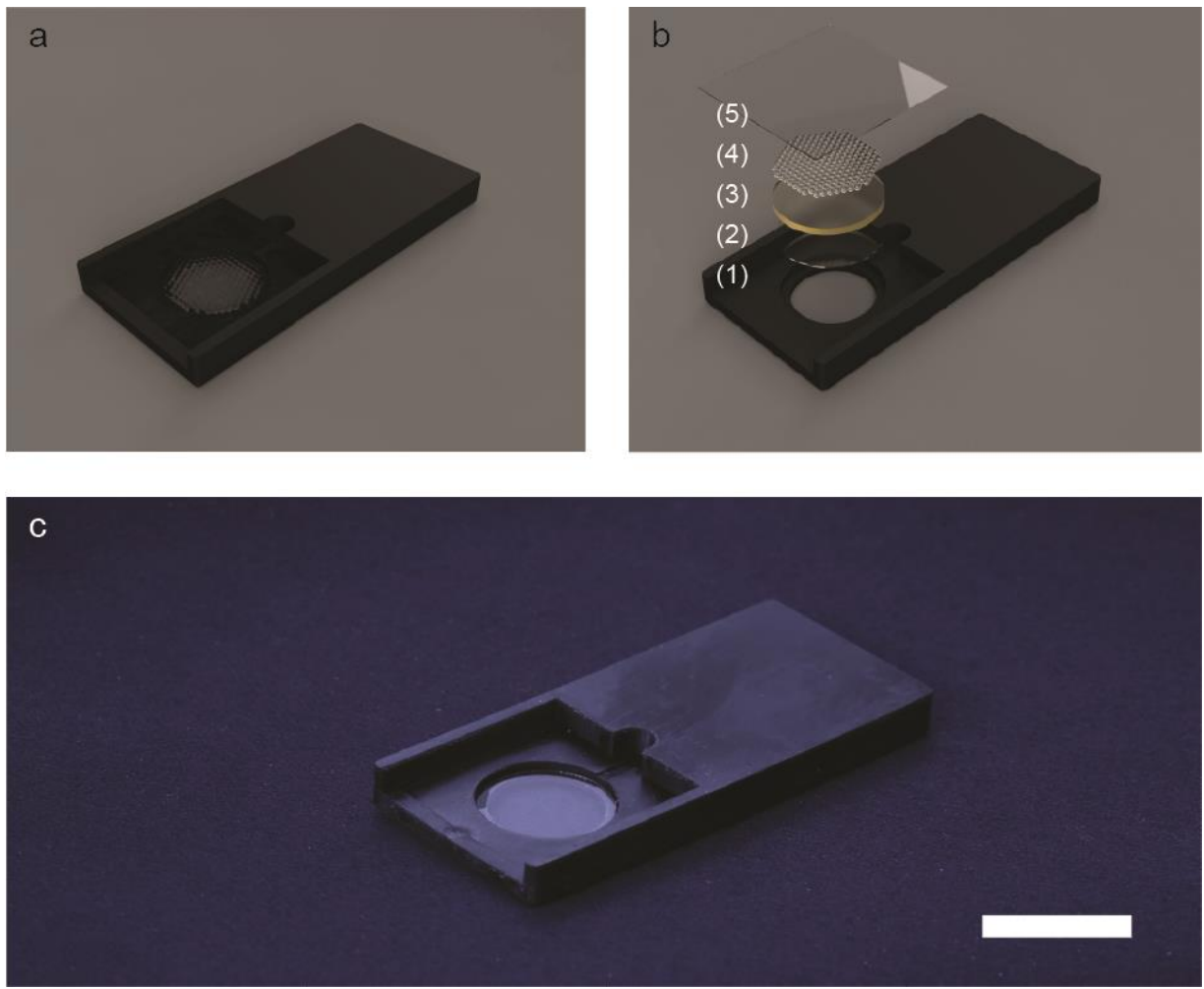

**Figure S1.** 3D printed slide used to contain the active microlens system. (a-b) Renderings of the sample holder in the (a) assembled and (b) exploded views. The rendering shown in (b) of the exploded view indicates (1) the 3D printed sample holder, (2) the first glass cover slip, (3) the layer of oil, (4) the layer of monodisperse droplets, and (5) the second glass cover slip. (c) Photograph of the 3D printed substrate. Scale bar is 2 cm.

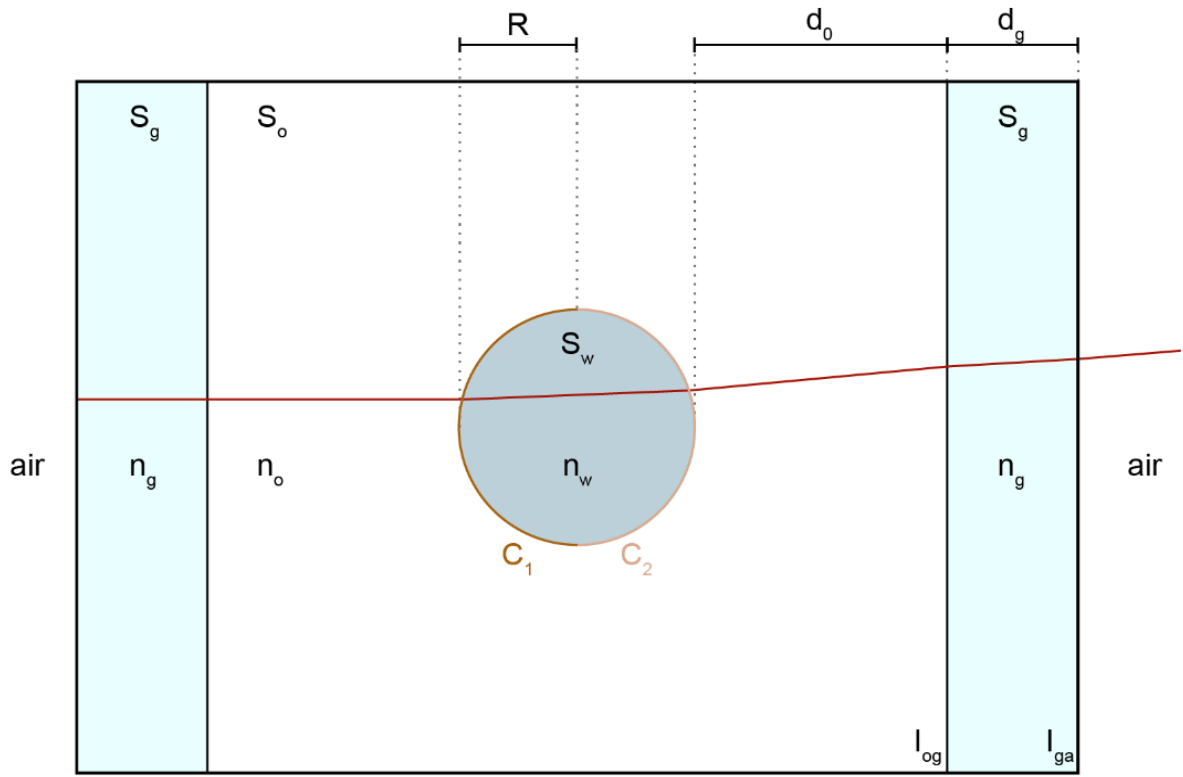

**Figure S2.** Schematic illustrating the different optical elements used for the analytical ray tracing model, including two glass slides ($S_g$) of refractive index $n_g$, the oil domain ($S_o$) with refractive index $n_o$ and the water droplet ($S_w$) with refractive index $n_w$. The interfaces between the optical elements are indicated by $C_1$, $C_2$, $I_{og}$ and $I_{ga}$. $R$, $d_o$ and $d_g$ represent the lengths and distances from the experimental system.

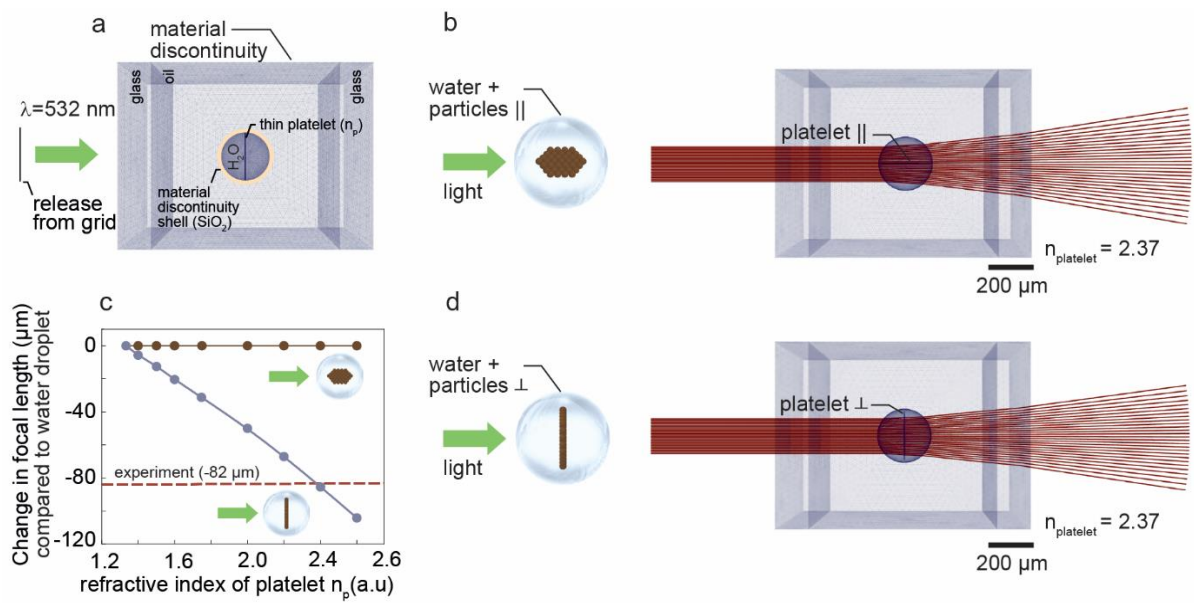

**Figure S3.** (a) Illustration of the three-dimensional ray optics model including boundary and ray release conditions. (b) Change in focal length compared to a water droplet ($n_{H_2O} = 1.333$) as function of platelet (cluster) refractive index ($n_p$) for two different orientations of a particle-made platelet (thickness of 4 monolayers). If the platelet is oriented parallel to the propagating rays, the focal length is independent of the platelet's refractive index. If the platelet is oriented perpendicular to the light rays, the focal length

decreases with increasing refractive index. Panels (c)-(d) display the ray trajectories ($\lambda = 532\ nm$) for platelet index of $n_p = 2.37$ for parallel or perpendicular alignment.

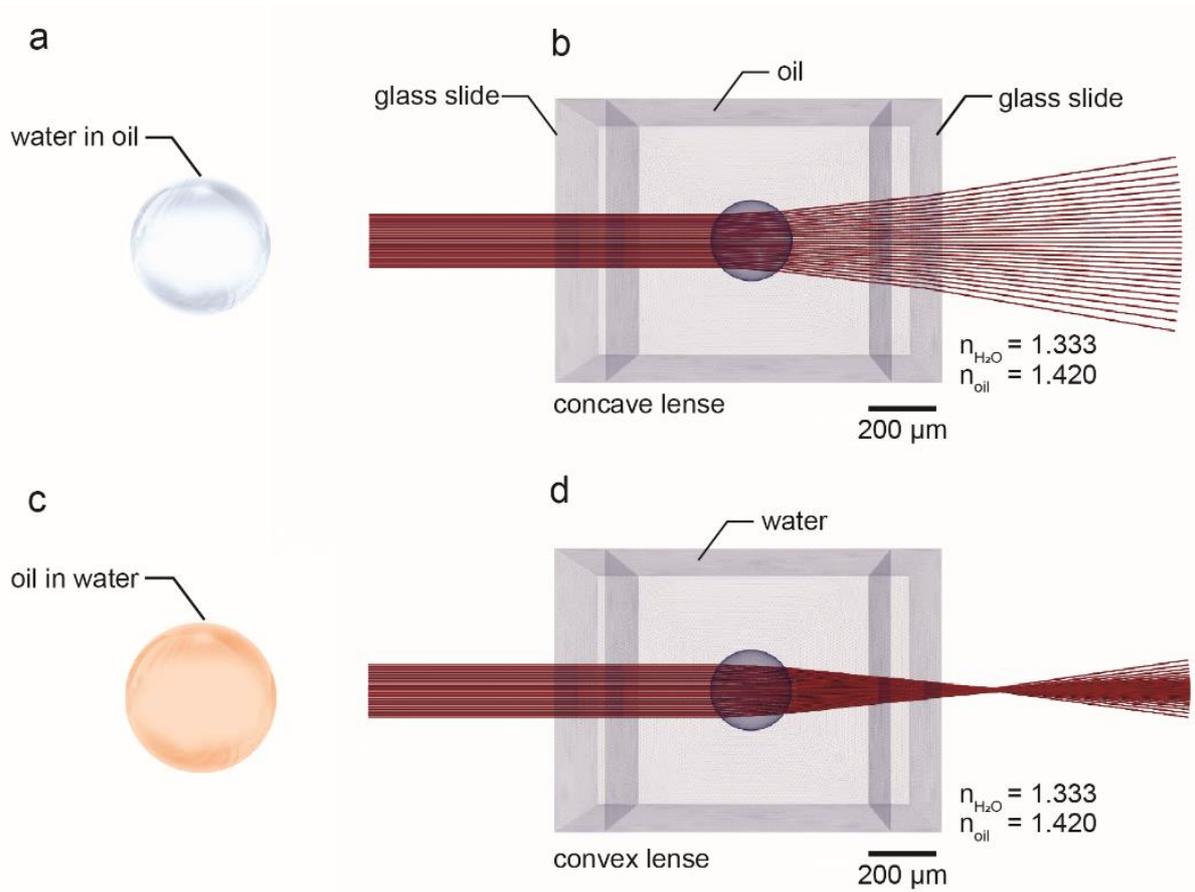

**Figure S4.** (a-b) Ray trajectories ($\lambda = 532\ nm$) for a water droplet with index of $n_{H_2O} = 1.333$ in oil ($n_{oil} = 1.420$). The microlens acts as a concave lens with negative focal length and negative focal power. (c-d) Ray trajectories ($\lambda = 532\ nm$) for an oil droplet with index of $n_{oil} = 1.42$ in water $n_{H_2O} = 1.333$. The microlens acts as a convex lens with positive focal length and positive focal power.

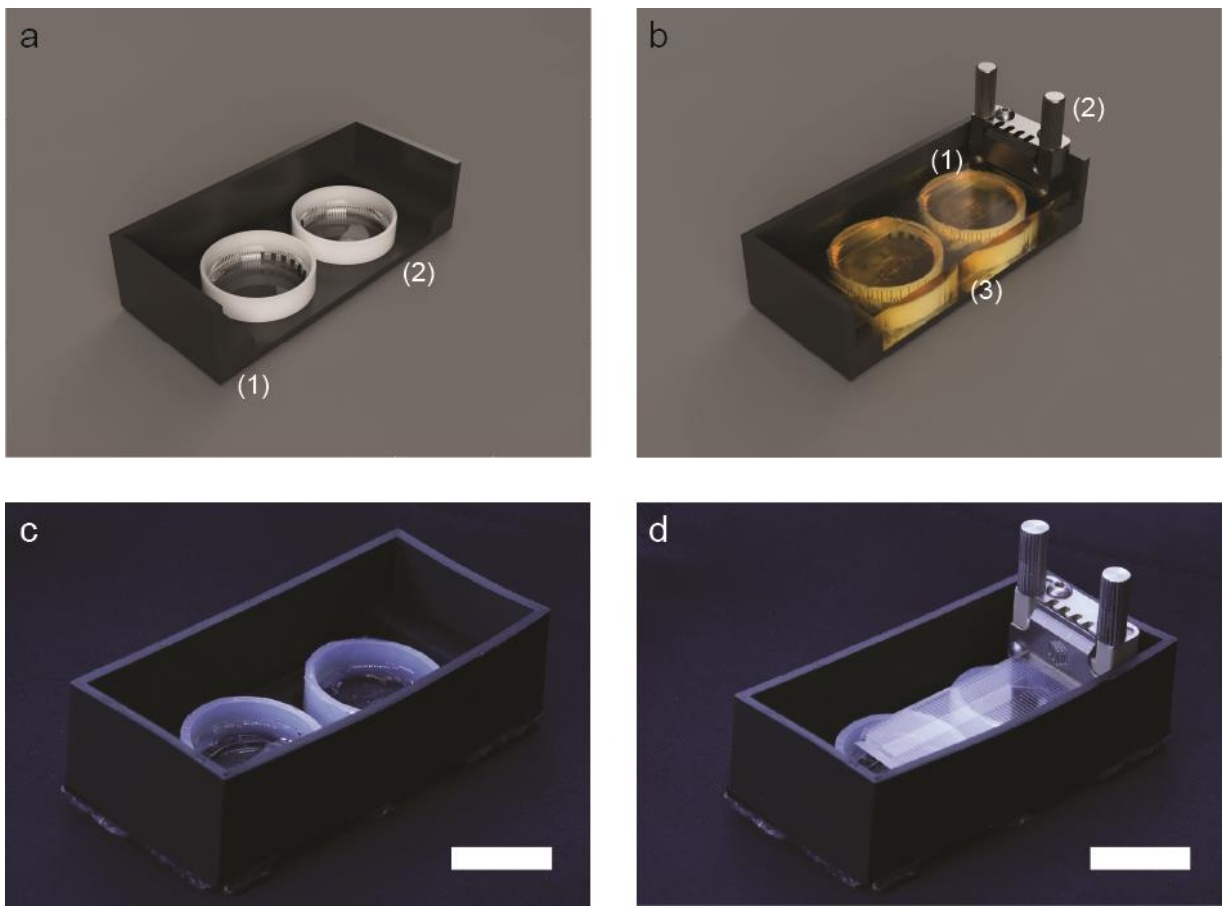

**Figure S5.** 3D printed emulsion container used to manufacture the active shutter elements. (a) Rendering of the (1) 3D printed container showing (2) two PTFE tubes partially filled with poly(dimethylsiloxane) (PDMS, Sylgard 184, Dow Corning). (b) Rendering of the printed holder with (1) the microfluidic device attached to (2) the Dolomite connector immersed in (3) the monomer mixture. (c-d) Photographs of the actual container alone and with the microfluidic device. Scale bar is 2 cm.

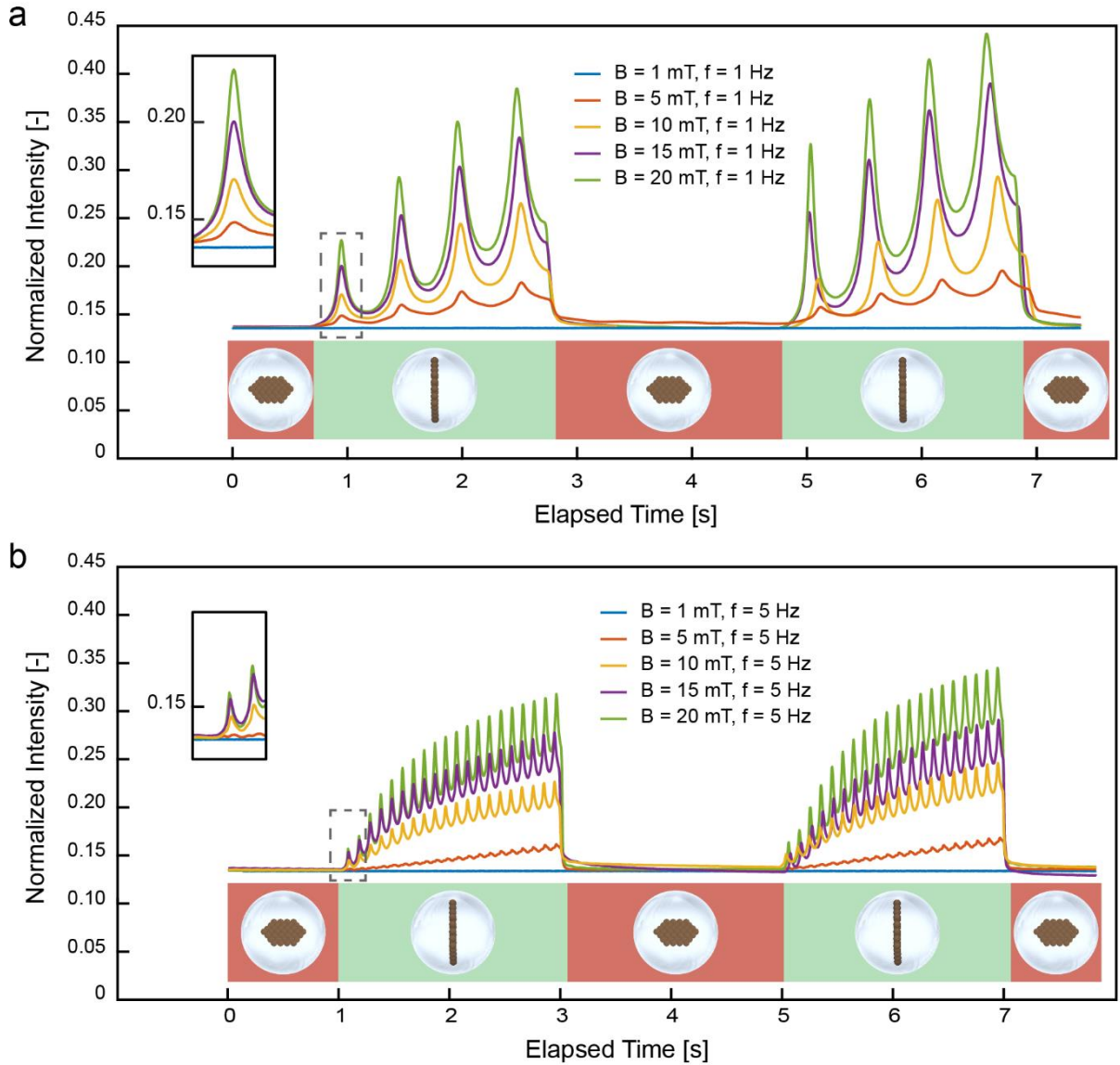

**Figure S6.** Droplet-based optical shutter controlled by magnetic fields. Intensity of light transmitted across the droplet array when the shutter is subjected to consecutive ON/OFF cycles. The ON state is triggered when the magnetic field is rotated within the xy-plane, whereas the OFF state is achieved when the applied field rotates within the yz-plane. The experiment data depicts the effect of the magnetic flux density *B* at a fixed frequency of (a) 1 Hz and (b) 5 Hz.